# Un Modelo Ontológico para el Gobierno Electrónico

Carlos Roberto Brys[1], José F. Aldana-Montes[2], David Luis La Red Martinez[3]

[1] Departamento de Informática, Facultad de Ciencias Económicas,
Universidad Nacional de Misiones,
Av. Fernando Llamosas Km 7,5. Campus UNaM.
N3304. Posadas. Misiones. Argentina
`brys@fce.unam.edu.ar`

[2] Departamento de Lenguajes y Ciencias de la Computación, Universidad de Málaga,
Boulevard Louis Pasteur 35. Campus de Teatinos.
29071 Málaga, España
`jfam@lcc.uma.es`

[3] Departamento de Informática, Facultad de Ciencias Exactas y Naturales y Agrimensura,
Universidad Nacional del Nordeste,
9 de Julio 1449
C3400. Corrientes. Argentina
`lrmdavid@exa.unne.edu.ar`

**Resumen.** La toma de decisiones muchas veces requiere información que debe suministrarse con el formato de datos enriquecidos. Atender apropiadamente estas nuevas exigencias hace necesario que las agencias gubernamentales orquesten grandes cantidades de información de diversas fuentes y formatos, para ser suministrados eficientemente a través de los dispositivos habitualmente utilizados por las personas, tales como computadoras, netbooks, tablets y smartphones.

Para superar estos problemas, se propone un modelo para la representación conceptual de las unidades de organización del Estado, vistos como entidades georeferenciadas del Gobierno Electrónico, basado en ontologías diseñadas bajo los principios de los Datos Abiertos Vinculados, lo que permite la extracción automática de información a través de las máquinas, que apoya al proceso de tomas de decisiones gubernamentales y da a los ciudadanos un acceso integral para encontrar y hacer trámites a través de las tecnologías móviles.

**Palabras clave:** Web de datos, Gobierno Electrónico, Administración Pública, Web Semántica, Datos Vinculados, Datos Abiertos del Gobierno, Extracción Automática de Datos

# 1 Introducción

La ventaja de expresar la estructura organizativa del Estado como una ontología gobierno electrónico, es que se puede construir un modelo de información que permita la exploración de datos en función de los elementos que representan las asociaciones entre los objetos, las propiedades de los elementos y formalmente describir la semántica de la clases y propiedades utilizadas en relación de dependencia, temporal y espacial; facilitando así que se pueda realizar un razonamiento automatizado, la búsqueda semántica y conceptual, y proporcionar servicios a los sistemas de apoyo a las decisiones.

El modelo presentado, incluye el diseño de una ontología para el Gobierno Electrónico de la Provincia de Misiones en Argentina, utilizando el Lenguaje de Ontologías Web (OWL)[1] como resultado de un proyecto de investigación en la Universidad Nacional de Misiones con el gobierno provincial.

# 2. Trabajos Relacionados

En 2004 el proyecto OntoGov desarrolló una plataforma semánticamente enriquecida con que aliviaría la composición coherente, la reconfiguración y evolución de los servicios de administración electrónica. El objetivo del proyecto OntoGov fue definir una ontología genérica de alto nivel para el ciclo de vida de servicio de Gobierno Electrónico, que serviría de base para el diseño de ontologías de dominio de nivel inferior específicos a los servicios públicos; desarrollar una plataforma semánticamente enriquecida que permita a las administraciones públicas modelar la semántica y las formalidades de sus servicios de administración electrónica [16].

Al mismo tiempo, Vassilakis y Lepouras [17] propusieron una ontología para los servicios públicos del gobierno electrónico donde introdujeron una primera aproximación de la modelación de una ontología para el Gobierno Electrónico, sobre la base de la organización, la legislación, la responsabilidad administrativa y de servicios.

Con el fin de estructurar el campo de la administración electrónica, los términos y la vinculación de los proyectos a través del uso de las tecnologías semánticas, en el Instituto de Informática en la Empresa y Gobierno (IWV) en la Universidad de Linz, se creó el proyecto del Portal Inteligente de Gobierno Electrónico. El resultado de este trabajo es una ontología de conocimiento y un mapa de gobierno electrónico [12].

---

[1] El Lenguaje de Ontologías Web (OWL) es una familia de lenguajes de representación del conocimiento, para publicar y compartir datos usando ontologías en la WWW.

Las "Ontologías para el eGovernment" (oeGOV), un proyecto desarrollado por TopQuadrant y dirigido por Ralph Hodgson, fue un trabajo pionero en la creación de una ontología para el gobierno electrónico [9]. El punto de partida del proyecto fue un modelado de las agencias de Estados Unidos y su estructura. Como resultado de la obra, se crearon y publicaron en el sitio web *www.oegov.org* una serie de ontologías fundacionales para el gobierno electrónico de Estados Unidos.

Para resolver los problemas de representación, Lacasta-Miguel [11] amplió un modelo de ontologías en tres niveles: una ontología de nivel superior que define los tipos de datos y las relaciones generales independientes del contexto. Una ontología de dominio donde los conceptos y relaciones reutilizables se definen en el contexto de los modelos administrativos de diferentes países; y una ontología de aplicación, donde están representados los tipos específicos de las unidades administrativas de cada país, junto con casos específicos de las unidades existentes.

Más enfocados en los servicios de gobierno electrónico, Hreño [10] y Ouchetto [13] plantearon el acceso, la recuperación y la integración de servicios utilizando ontologías. Considerando que la terminología relacionada con el campo de la administración electrónica es variada, proponen dividir la ontología en las sub-ontologías (ontologías sectoriales). Para resolver la complejidad de la modelización conceptual en escenarios complejos basados en la interoperabilidad semántica [2] proponen un método mediante el uso de ontologías de dominio y donde las fuentes de información son bases de datos, documentos legales y las personas.

## 3. Marco Teórico

Vamos a utilizar el término "Estado" como definición de un concepto político que se refiere a una forma de organización social y políticamente soberana, formada por un grupo de instituciones. Estas instituciones se estructuran funcionalmente en unidades administrativas, que son los elementos básicos de las estructuras organizativas. En general, la organización de un Estado se distingue por: Funciones, Instituciones y Autoridades. El alto grado de diversidad y especialización de las unidades administrativas demandan un modelo coherente para facilitar su gestión y simplificar el uso [11].

### 3.1. Datos Abiertos Vinculados del Gobierno (LGOD)

Tim Berners Lee esbozó un conjunto de normas para la publicación de datos en la Web, de forma que todos los datos publicados se convierten en parte de un espacio único de datos globales [1]. Estos son conocidos como los "Principios de los Datos Vinculados". Datos Vinculados (Linked Data) es un modelo estándar para el intercambio de datos en la web. Este término se utiliza para describir una práctica recomendada para exponer y compartir piezas de conexión de datos, información y conocimiento sobre la Web Semántica utilizando URI y RDF.

La utilización de las tecnologías semánticas es imprescindible para un Gobierno Electrónico con LGOD. Con LGOD, los ciudadanos pueden utilizar la web para vincular datos y utilizarlos aunque no estuvieran vinculados con anterioridad, generando aplicaciones acordes a sus necesidades [3].

### 3.2. Ubicación Espacial de los Datos: GeoDatos

Según [6], del 60% al 80% de las decisiones que afectan a los ciudadanos se relacionan con la información geográfica, la cual es cada vez más importante es aspectos vitales como el transporte, la energía, la agricultura, la protección del medio ambiente, la silvicultura, las regulaciones para el uso de la tierra, el desarrollo planificado, las TIC, la cultura, la educación, los seguros, la defensa nacional, la atención de la salud, la seguridad interna, la prevención de desastres, la defensa civil, y la provisión de servicios públicos.

LinkedGeoData[2] añade una dimensión espacial a la Web de Datos. Permite levantar datos de OpenStreetMap[3] en la infraestructura de la Web Semántica, y los hace accesibles como una base de conocimiento RDF según los Principios de los Datos Vinculados. Esto simplifica las tareas de integración de información y de agregación que requieren un amplio conocimiento de fondo relacionado con el ámbito espacial [15].

## 4. Un Nuevo Modelo de Ontología de Gobierno Electrónico (OGE)

Según Gruber [7], las ontologías son una manera formal y explícita para definir una conceptualización para el intercambio de conocimientos. La idea fundamental detrás de estas tecnologías es lograr que las computadoras puedan comprender por sí solas los datos, minimizando la participación humana en dicho proceso.

---

[2] LinkedGeoData http://linkedgeodata.org/
[3] OpenStreetMap http://www.openstreetmap.org/

Con el fin de especificar la OGE para representar al Estado y su estructura organizativa partimos del modelo de tres capas propuesto por Guarino [8] y revisado por Lacasta-Miguel [11], y lo extendemos añadiendo una nueva capa de ontologías bajo los principios de los Datos Abiertos Vinculados, como se muestra en la Figura 1.

El modelo de ontología extendido tiene:

• Una ontología de alto nivel que define los conceptos de Estado, los Poderes, el marco legal, los conceptos básicos, tipos de datos y las relaciones generales, independiente del contexto.

• Una ontología de dominio que define la estructura organizativa y describe en detalle las unidades administrativas específicas, sus jerarquías, dependencias y relaciones.

• Otras ontologías externas y sus relaciones de vinculación a través de los Datos Abiertos Vinculados que enriquecen los datos de los individos instanciados y provee la información geoespacial.

• Y varias ontologías de aplicación de los trámites, servicios, y sus instancias.

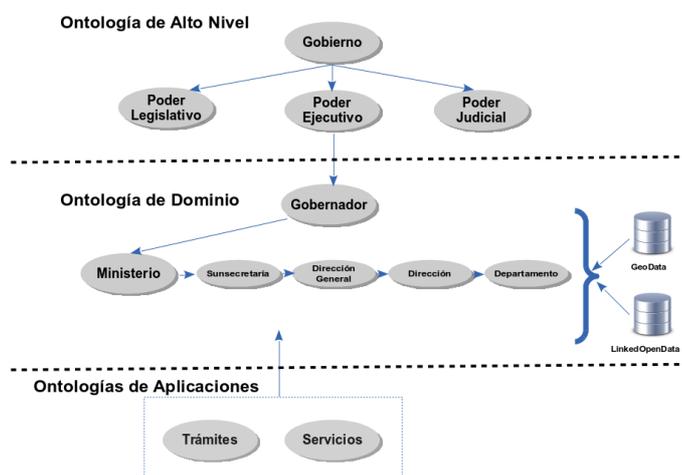

**Fig. 1.** Modelo de la Ontología de Gobierno Electrónico.

Para construir la OGE, utilizamos la metodología Methontology [4], [5], que fue creada en el Laboratorio de Inteligencia Artificial de la Universidad Politécnica de Madrid. La razón de esta elección es el fuerte apoyo de herramientas de software, la independencia de la plataforma, es recomendada por la Fundación para Agentes

Físicos Inteligentes (FIPA) para el desarrollo de ontologías, ha sido probada en varios proyectos a gran escala y se ha aplicado con éxito en el desarrollo de ontologías para la gestión del conocimiento de gobierno abierto [14], [2].

Por lo tanto, siguiendo las pautas metodológicas de Methontology, para la *especificación*, se definió el alcance y la granularidad. Para el proceso de *conceptualización*, se relevó la estructura organizacional y las normas legales que la sustentan. Estos datos actualmente se registran en formularios de papel y organigramas. Los diferentes conceptos que se estudiaron fueron enumerados, a continuación, se agruparon por similitud y utilidad. Para el mapeo de los organigramas del árbol de la estructura administrativa utilizamos la aplicación CmapTools COE. Para la etapa de *implementación*, los organigramas representados en los mapas conceptuales fueron exportados al Lenguaje de Ontologías Web (OWL) y luego editadas usando el editor OWL Protégé de la Universidad de Standford para añadir las clases, atributos, relaciones y datos de geolocalización.

En el nivel superior, definimos las super clases relacionadas con conceptos generales, como Estado, el Gobierno, los Poderes, el marco legal y el territorio. Luego, para el nivel de dominio específico, definimos las clases de gobierno electrónico para la estructura administrativa y de los puestos, la gobernación y su estructura ministerial, la infraestructura (edificios), las oficinas de atnción al público y los agentes. En esta etapa hemos modelado la organización administrativa de acuerdo con la recomendación de la W3C para ontologías de organización [18]. Esta ontología describe a la gobernación, la vicegobernación, 10 ministerios, 37 subsecretarías, 68 direcciones generales, 114 direcciones, y 326 departamentos.

Finalmente en el nivel de aplicación definimos las clases para los trámites y servicios para los ciudadanos. Para cada clase, buscamos sus instancias, las incorporamos a la ontología específica y las mapeamos con otras si fuera necesario.

Esta relación entre el Estado, el gobierno y sus divisiones se puede representar en lógica descriptiva con la propiedad *has*, el subgrupo, la cuantificación existencial y Símbolos de la unión de la siguiente manera:

Estado $\subseteq \exists$ hasPower.(Ejecutivo $\cup$ Legislativo $\cup$ Judicial)
Ejecutivo $\subseteq \exists$ hasDivision. (Gobernador $\cup$ (Ministerio $\cup$ ... $\cup$ Ministerio)
Ministerio $\subseteq$ Subsecretarías $\subseteq$ Direcciones Generales $\subseteq$ Direcciones $\subseteq$ Departamentos

Además de la organización administrativa, se necesitan representar a los otros elementos del Estado. Para ello, desarrollamos las ontologías que describen el marco jurídico, el territorio (para lugares geolocalizados), la infraestructura (para oficinas administrativas) y la estructura de los puestos administrativos (las personas).

### 4.1. Mapeo de Ontologías

Un modelo de ontología taxonómica no puede representar satisfactoriamente la complejidad del Estado, entonces es necesario ampliar el dominio de la OGE vinculándola con otras fuentes de información externa y abierta. Para esto, son reutilizadas otras ontologías por ejemplo OSMonto: una ontología de las etiquetas de OpenStreetMap, donde cada organización administrativa individual en la OGE está representada en OpenStreetMap por un "nodo" o una "via". Estos nodos tienen las "etiquetas" definidas en la ontología OSMonto que coinciden con algunos atributos de los individuos en la ontología OGE. También relacionamos la OGE con otras fuentes de datos como FOAF, DBpedia, GeoNames y LinkedGeoData, como se muestra en la Figura 2.

En un primer paso, nos enfocamos sólo en las oficinas administrativas que están abiertas al público y ofrecen un servicio del gobierno. Cada individuo está vinculado a su nodo geolocalizado en OpenStreetMap, y GeoNames. Esto permite la integración con un amplio espectro de fuentes de datos abiertos locales y externas.

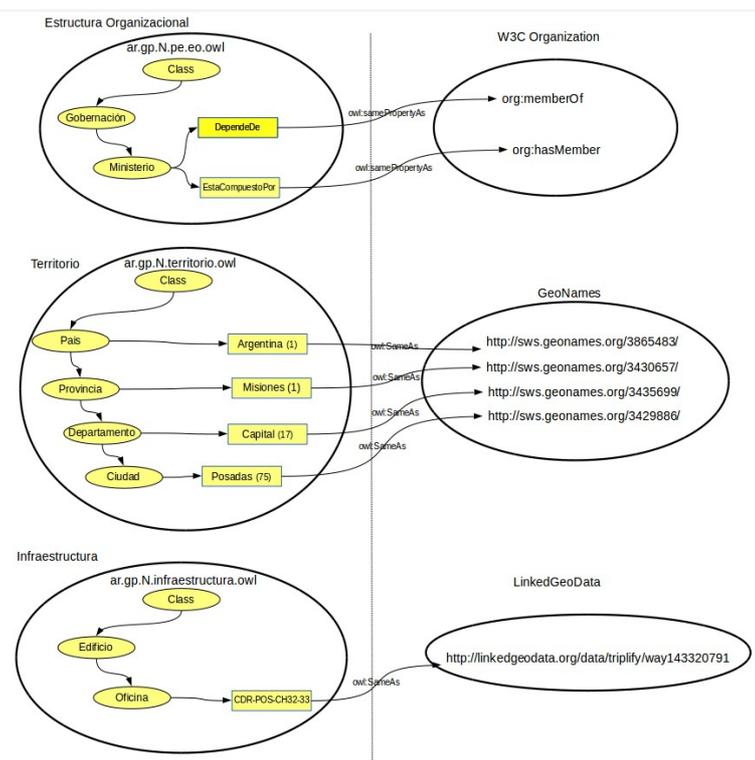

**Fig. 2.** Mapeo entre de la Ontología de Gobierno Electrónico y Datos Abiertos Vinculados.

Como un caso de uso, planteamos una situación donde un ciudadano desea obtener un nuevo DNI. La consulta a la ontología se puede representar en un grafo, tal como se muestra la Figura 3, que representa gráficamente un ejemplo de las instancias interrelacionadas de "Nuevo_DNI" y "Posadas".

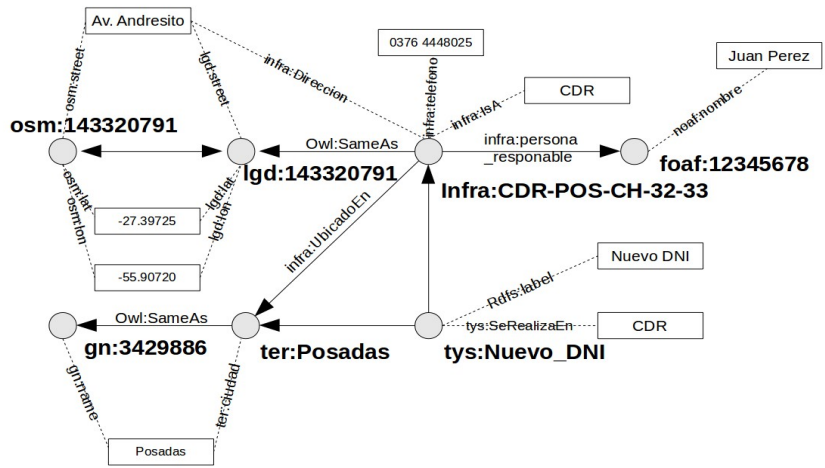

**Fig. 3.** Instancias enlazadas para Nuevo_DNI y Posadas.

Definiciones: V es un conjunto de vértices que son las etiquetas de los conjuntos de datos que tienen las instancias *SameAs* enlazados, E ⊆ V × V es un conjunto de aristas *sameAs*, e I es un conjunto de URIs de las instancias sameAs interrelacionadas.

GDNI = (V, E, I), donde

V = {S, T, O, L, I, G, F},

E = {(S, T), (S, I), (I, F), (I, L), (L, O), (T, G)},

I = {TYS:Nuevo_DNI, ter:Posadas, osm:143320791, lgd:143320791, infra:CDR-POS-CH-32-33, gn:3429886, foaf:12345678.

S, T, O, L, I, G y F representan las etiquetas de los onjuntos de datos de Trámites y Servicios, Territorio, OSM, LinkedGeoData, Infraestructura, GeoNames y Personas respectivamente.

## 5. Conclusiones y Trabajos Futuros

Hemos propuesto un enfoque original que replantea la forma en que la literatura existente define la forma de la prestación de servicios públicos, en los que los ciudadanos tienen que saber dónde y cómo hacer sus trámites. Con este modelo, los servicios públicos son los que encuentran a los ciudadanos y lo entregan dondequiera que ellos estén, en sus propios dispositivos móviles.

También llegamos a la conclusión de que el actual modelo es adecuado como un marco de apoyo para la integración y la interoperabilidad de los servicios públicos prestados por las oficinas de administración distribuidas geográficamente.

En conclusión, el modelo propuesto es un paso exitoso en la evolución del gobierno electrónico a un nivel semántico, contribuye a la integración e interoperabilidad de los procesos que van más allá de las fronteras geográficas y los estados administrativos, y alcanzar la meta más importante del gobierno electrónico: prestar servicios más eficientes para los ciudadanos, ahorrando su tiempo y dinero.

**Referencias**

1. Berners-Lee T. (2006). Linked Data. http://www.w3.org/DesignIssues/LinkedData.html
2. Brusa, G., Caliusco, M. L., & Chiotti, O. (2013). Gestión del Conocimiento en el Gobierno Abierto: Ontologías de Dominio. In JAIIO (Ed.), 42 Jorandas Argentinas de Informática. 7mo Simposio Argentino De Informatica En El Estado (pp. 8–22). Córdoba. Retrieved from http://42jaiio.sadio.org.ar/proceedings/simposios/Trabajos/SIE/12.pdf
3. EGW3C (2008) eGovernment at W3C. Use Case: Open Government: Linked Open Data. http://www.w3.org/egov/wiki/Use_Case_8_-_Linked_Open_Government
4. Fernández-López, M., Gómez-Pérez, A., Juristo, N. (1997): Methontology: from ontological art towards ontological engineering. In Proc. Symposium on Ontological Engineering of AAAI. (33-40)
5. Gómez-Pérez A., Fernández-López M., Corcho O., (2004). Ontological Engineering, Editorial: Springer Verlag GmH & Co. ISBN: 1-85233-551-3
6. GPSC The Geoportal of the Swiss Confederation. (2011) http://www.geo.admin.ch/internet/geoportal/en/home.htm
7. Gruber, T. (1993). A translation approach to portable ontology specifications. Knowledge Acquisition 199–220.
8. Guarino N. (1998). Formal Ontologies and Information Systems. Proceedings of FOIS'98, 3-15. Trento, Italy
9. Hodgson R., Allemang D., (2006), Semantic Technology For e-Government, Semantic Web and Beyond Volume 3, 2006, pp 283-303, Springer US, ISBN 978-0-387-30239-3.
10. Hreño, J., Bednár, P., Furdík, K., & Sabol, T. (2011). Integration of government services using semantic technologies. Journal of Theoretical and …, 6(1), 143–154. doi:10.4067/S0718-18762011000100010

11. Lacasta-Miguel, J., López-Pellicer, F.J., Floristán-Jusué, J., Nogueras-Iso, J., Zarazaga-Soria, F.J.(2006): Unidades administrativas, una perspectiva ontológica. Avances en las Infraestructuras de Datos Espaciales. Treballs dínformática i tecnologia. Castelló de la Plana: Universidad Jaime I de Castellón. (85-94). ISBN 84-8021-590-9.
12. Orthofer G. and Wimmer M., (2006), An Ontology for eGovernment: Linking the Scientific Model with Concrete Projects, Semantic Web Meets eGovernment, Papers from the 2006 AAAI Spring Symposium: Semantic Web Meets eGovernment, pp. 96-98.
13. Ouchetto, H., Ouchetto, O., & Roudiès, O. (2012). Ontology-oriented e-gov services retrieval. IJCSI International Journal of Computer Science, 9, 99–107.
14. Sabucedo L., Rifon L., (2006) Semantic Service Oriented Architectures for eGovernment Platforms. American Association for Artificial Intelligence. AAAI. In Proc. 2006 AAAI Spring Symposium.
15. Stadler C., Lehmann J., Höffner K., Auer S., (2011) LinkedGeoData: A Core for a Web of Spatial Open Data. Semantic Web -Interoperability, Usability, Applicability an IOS Press Journal, 1570-0844/0-1900
16. Tambouris E., Gorilas S., Kavadias G., Apostolou D., Abecker A., Stojanovic L., Mentzas G., (2004), Ontology-enabled E-government Service Configuration-the OntoGov Approach. Springer Berlin Heidelberg. Knowledge Management in Electronic Government. Lecture Notes in Computer Science Volume 3035, 2004, pp 122-127.
17. Vassilakis, C., & Lepouras, G. (2006). Ontology for e-Government Public Services. Encyclopedia of E-Commerce, E-Government, and Mobile Commerce (pp. 865-870).
18. (W3C) The World Wide Web Consortium (2013). Recommendation for "The Organization Ontology". Available a. http://www.w3.org/TR/vocab-org/